\documentclass[twocolumn]{aastex631}

\usepackage{academicons}
\usepackage{amsmath}
\usepackage{hyperref}
\usepackage{natbib}
\usepackage{xcolor}
\usepackage{xspace}
\usepackage{subfigure}
\usepackage{threeparttable}
\usepackage{makecell}

\shortauthors{Van Zandt et al.}

\graphicspath{{./}{Figures/}}
\newcommand{\MA}{\ensuremath{1.26 \pm 0.01}}
\newcommand{\FeHA}{\ensuremath{0.20 \pm 0.07}}
\newcommand{\dmu}{\ensuremath{13.8 \pm 0.2}}

\newcommand{\MBsolar}{\ensuremath{0.396 \pm 0.002}}
\newcommand{\MBjup}{\ensuremath{415 \pm 2}}
\newcommand{\PB}{\ensuremath{69.9 \pm 0.4}}
\newcommand{\eB}{\ensuremath{0.422 \pm 0.002}}

\newcommand{\aB}{\ensuremath{20.07 \pm 0.06}}
\newcommand{\iB}{\ensuremath{119.8 \pm 0.1}}
\newcommand{\OB}{\ensuremath{17.2 \pm 0.2}} 
\newcommand{\oB}{\ensuremath{341.9 \pm 0.2}} 
\newcommand{\tpB}{\ensuremath{2473832^{+245}_{-236}}}

\newcommand{\MAbsini}{\ensuremath{1.62^{+0.04}_{-0.04}}}
\newcommand{\PAb}{\ensuremath{2.475 \pm 0.002}}
\newcommand{\eAb}{\ensuremath{0.07 ^{+0.03}_{-0.03}}}

\newcommand{\aAb}{\ensuremath{1.978 \pm 0.007}}
\newcommand{\iAb}{\ensuremath{90 \pm 40}}
\newcommand{\OAb}{\ensuremath{0^{+100}_{-100}}} 
\newcommand{\oAb}{\ensuremath{295 \pm 20}} 
\newcommand{\tpAb}{\ensuremath{2455980^{+46}_{-50}}}

\newcommand{\Me}{\ensuremath{M_{\oplus}}\xspace}

\newcommand{\Msun}{\ensuremath{M_{\odot}}\xspace}
\newcommand{\Mjup}{\ensuremath{M_{\mathrm{Jup}}}\xspace}

\begin{document}

\title{The $\gamma$ Cephei System: Updated Orbits, Dynamical Architecture, and Limits on Additional Companions}

\correspondingauthor{Judah Van Zandt}
\email{judahvz@ucsb.edu}

\author[0000-0002-4290-6826]{Judah Van Zandt}
\affiliation{Department of Physics, University of California, Santa Barbara, CA 93106, USA}

\author[0000-0003-2649-2288]{Brendan P. Bowler}
\affiliation{Department of Physics, University of California, Santa Barbara, CA 93106, USA}

\author[0000-0002-7714-6310]{Michael Endl}
\affiliation{Department of Astronomy, The University of Texas at Austin, Austin, TX 78712, USA}
\affiliation{McDonald Observatory, The University of Texas at Austin, Fort Davis, TX 79734, USA}
\affiliation{Center for Planetary Systems Habitability, The University of Texas at Austin, Austin, TX 78712, USA}

\author[0000-0001-9662-3496]{William D. Cochran}
\affiliation{Department of Astronomy, The University of Texas at Austin, Austin, TX 78712, USA}
\affiliation{McDonald Observatory, The University of Texas at Austin, Fort Davis, TX 79734, USA}
\affiliation{Center for Planetary Systems Habitability, The University of Texas at Austin, Austin, TX 78712, USA}

\author[0009-0006-7054-0100]{Phillip MacQueen}
\affiliation{McDonald Observatory, The University of Texas at Austin, Fort Davis, TX 79734, USA}

\author{Artie Hatzes}
\affiliation{Th{\"u}ringer Landessternwarte Tautenburg, Sternwarte 5, 07778, Tautenburg, Germany}

\author[0000-0002-5286-0251]{Guillermo Torres}
\affiliation{Center for Astrophysics $\vert$ Harvard \& Smithsonian, 60 Garden St., Cambridge MA 02138, USA}

\author[0000-0001-9911-7388]{David W. Latham}
\affiliation{Center for Astrophysics $\vert$ Harvard \& Smithsonian, 60 Garden St., Cambridge MA 02138, USA}

\author[0000-0001-8638-0320]{Andrew W. Howard}
\affiliation{Department of Astronomy, California Institute of Technology, Pasadena, CA 91125, USA}

\author[0000-0003-3504-5316]{Benjamin Fulton}
\affiliation{NASA Exoplanet Science Institute/Caltech-IPAC, MC 314-6, 1200 E. California Blvd., Pasadena, CA 91125, USA}

\author[0000-0002-0531-1073]{Howard Isaacson}
\affiliation{{Department of Astronomy, University of California Berkeley, Berkeley CA 94720, USA}}
\affiliation{Centre for Astrophysics, University of Southern Queensland, Toowoomba, QLD, Australia}

\author[0000-0003-2232-7664]{Michael C. Liu}
\affiliation{Institute for Astronomy, University of Hawai'i, 2680 Woodlawn Drive, Honolulu, HI 96822, USA}

\author[0000-0001-7062-815X]{Samuel A. U. Walker}
\affiliation{Institute for Astronomy, University of Hawai'i, 2680 Woodlawn Drive, Honolulu, HI 96822, USA}

\author[0000-0002-6618-1137]{Jerry W. Xuan}
\altaffiliation{51 Pegasi b Fellow}
\affiliation{Department of Earth, Planetary, and Space Science, 595 Charles E. Young Dr E, University of California, Los Angeles, CA 90095, USA}

\author[0000-0002-2696-2406]{Jingwen Zhang}
\affiliation{Department of Physics, University of California, Santa Barbara, CA 93106, USA}

\author{Rebeca E. Soto Armendariz}
\affiliation{Department of Astronomy, The University of Texas at Austin, Austin, TX 78712, USA}

\author[0000-0003-2646-3727]{Lauren I. Biddle}
\affiliation{Department of Astrophysical Sciences, Princeton University, Princeton, NJ 08540, USA}

\author[0000-0003-4557-414X]{Kyle Franson}
\altaffiliation{NHFP Sagan Fellow}
\affiliation{Department of Astronomy $\&$ Astrophysics, University of California, Santa Cruz, CA 95064, USA}

\author[0000-0003-4006-102X]{Lillian Jiang}
\affiliation{Department of Physics, University of California, Santa Barbara, CA 93106, USA}

\author[0000-0003-4022-6234]{Marvin Morgan}
\affiliation{Department of Physics, University of California, Santa Barbara, CA 93106, USA}

\author[0000-0001-6532-6755]{Quang H. Tran}
\altaffiliation{51 Pegasi b Fellow}
\affiliation{Department of Astronomy, Yale University, New Haven, CT 06511, USA}

\begin{abstract}
The $\gamma$ Cephei system hosts one of the first exoplanets discovered and is orbited by one of the closest known stellar companions to a planet-hosting star. In this work, we derive updated orbital fits for $\gamma$ Cep AB, the stellar binary, and Ab, the planet, by combining literature data with \textit{Hipparcos-Gaia} astrometry, new radial velocities (RVs), and adaptive optics imaging. We acquired 328 RVs of $\gamma$ Cep A with Keck/HIRES, AFP/Levy, McDonald/Tull, and Whipple/TRES, and eight adaptive optics imaging epochs with Keck/NIRC2, including the earliest spatially resolved image of $\gamma$ Cep B in 2003. These observations extend the precision RV baseline of $\gamma$ Cep to 45 years and the direct imaging baseline to 23 years, improving the inferred orbital parameter precision by a factor of 2--10 compared to previous work. For $\gamma$ Cep B, we derive a semi-major axis of $a_B=\aB$ AU, a mass of $M_B=\MBjup \, \Mjup$ ($\MBsolar~\Msun$), an eccentricity of $e_B=\eB$, and an inclination of $i_B=\iB^{\circ}$. For $\gamma$ Cep Ab, we find a semi-major axis of $a_{Ab}=\aAb$ AU, a minimum mass of $M_{Ab} \sin i = \MAbsini~\Mjup$, and an eccentricity of $e_{Ab}=\eAb$. The system is dynamically compact and its orbital architecture is consistent with von Zeipel-Lidov-Kozai (vZLK) oscillations, although the planet's orbital inclination remains insufficiently constrained to determine whether the vZLK mechanism is active. Using the RV residuals and dynamical constraints, we rule out additional Jovian companions between 2.5--20 AU, and companions more massive than Neptune for $a<1$ AU, both at $>90\%$ confidence. The absence of additional giant planets over a broad range of orbital separations is consistent with a dynamically sculpted system in which the close stellar companion has strongly limited the formation or long-term survival of other long-period giant companions.
\end{abstract}


\section{Introduction}
\label{sec:intro}

$\gamma$ Cephei (also known as $\gamma$ Cep, HD 222404, HIP 116727, and HR 8974) is a bright, nearby, evolved star ascending the lower red giant branch ($G$=2.9 mag, $d$=13.8 pc, spectral type K1 III-IV; \citealt{GaiaEDR32021_content, Keenan1989}). Due to its brightness, $\gamma$ Cep has a rich history of astrometric and radial velocity observations reaching back over a century \citep{FrostAdams1903, Slipher1905, Kustner1908}, and was among the first stars targeted for a precision radial velocity (RV) search for planetary companions by \cite{Campbell1988}, who determined $\gamma$ Cep to be a single-lined spectroscopic binary from over six years of observation. They also noted a second signal superimposed over the first, with a 2.7-year period and an amplitude of 25 m/s, corresponding to the first-ever candidate extrasolar planetary companion orbiting a Sun-like star, with a minimum mass of $M\sin i \sim \, 1.7 \, \Mjup$.

The same group \citep{Walker1992} later cast doubt on a planetary explanation for the second signal by extending $\gamma$ Cep's precision RV baseline to 11 years and matching the revised 2.52-year period to 2.48-year variability in the emission in the core of the star's Ca II absorption line at 866.2 nm \citep{Andretta2005}, suggesting that the periodicity may be linked to activity instead of gravitational reflex motion from a planet. They also set a lower limit of $\sim$30 years on the orbital period of the stellar companion, $\gamma$ Cep B. 


The planetary nature of $\gamma$ Cep Ab was established a decade later by \cite{Hatzes03}, who reanalyzed the archival CFHT RVs together with $\sim$15 years of precise RVs from the McDonald Observatory 2.7m Harlan J. Smith Telescope coud\'{e} focus \citep{Tull1995, CochranHatzes2004}. They refined the primary star's activity cycle to 2.14 years and found that unlike the RV signal, the activity-driven variability reported by \cite{Walker1992} did not persist over $\gamma$ Cep's entire 20-year precision RV observing history, implying activity is unlikely to be responsible for the long-lived periodicity in the RVs. Their extended baseline also permitted an updated estimate of $\gamma$ Cep B's period ($P$=57 years) and eccentricity ($e$=0.36).

\cite{Torres07} improved orbital constraints for both companions by incorporating new RVs and astrometric measurements as well as evolutionary models to revise $\gamma$ Cep A's mass, a key source of uncertainty in past determinations of the binary orbital parameters. Ground-based and $Hipparcos$ \citep{Hipparcos1997} astrometry provided constraints on the binary inclination, resulting in a true mass estimate for $\gamma$ Cep B: $0.362 \pm 0.022 \, \Msun$, consistent with an M dwarf. Both \cite{Neuhauser07} and \cite{Mugrauer22} derived dynamical masses of $\gamma$ Cep B using literature RVs and new direct imaging observations, confirming its status as an M dwarf.

\cite{ReffertQuirrenbach2011} used the re-reduction of the $Hipparcos$ mission's astrometric data \citep{vanLeeuwen2007} to place more precise astrometric constraints on $\gamma$ Cep Ab. They fit only inclination and longitude of the ascending node, fixing the remaining orbital parameters to the RV-derived literature values available at that time. Their results indicated two degenerate solutions: $i_{Ab}=3.8^{\circ}-20^{\circ}$ and $\Omega_{Ab}=352.9^{\circ}-86^{\circ}$, and $i_{Ab}=166.6^{\circ}-174.8^{\circ}$ and $\Omega_{Ab}=330.2^{\circ}-25.0^{\circ}$. \cite{Benedict2018} later performed a similar fit by incorporating new Hubble Space Telescope Fine Guidance Sensor astrometry of $\gamma$ Cep and derived a consistent inclination ($169.5\pm1.1^{\circ}$), but a discrepant ascending node ($47\pm6^{\circ}$).

Most recently, building on similar studies by \cite{Stello2017} and \cite{Malla2020}, \cite{Knudstrup2023} conducted a detailed refit of $\gamma$ Cep A's stellar parameters --- including mass, radius, and age --- by studying asteroseismic oscillations using $TESS$ photometry and high-cadence RVs from the SONG network \citep{Grundahl2008, Grundahl2017}. They found a mass of $M_A=1.27^{+0.05}_{-0.07}~\Msun$, a radius of $R_A=4.74^{+0.07}_{-0.08}~R_{\odot}$, and an age of $\tau_A=5.7^{+0.8}_{-0.9}$ Gyr, identifying $\gamma$ Cep A as a ``retired" F star ascending the red giant branch (RGB). These revised parameters, together with a measurement of $\gamma$ Cep B's inclination adopted from \cite{Neuhauser07} and the dynamically inferred constraints for $\gamma$ Cep Ab's inclination from \cite{ReffertQuirrenbach2011}, propagated to improved estimates of the companion periods and true masses: $P_B = 58.0 \pm 0.1$ years, $M_B = 0.328^{+0.009}_{-0.012} \, \Msun$, $P_{Ab} = 913 \pm 3$ days, $M_{Ab} = 6.6^{+2.3}_{-2.8} \, \Mjup$. However, inconsistencies between the two measurements of $\gamma$ Cep Ab's inclination to date, by \cite{ReffertQuirrenbach2011} and \cite{Benedict2018}, have left both the planet inclination and the true mass uncertain.

In this work, we uniformly refit literature RVs, imaging, and both relative and absolute astrometry, together with new RVs measured with the Keck/HIRES \citep{Vogt1994}, APF/Levy \citep{Vogt2014}, Harlan J. Smith/Tull \citep{Tull1995}, and F.L. Whipple/TRES \citep{SzentgyorgyiFuresz2007} spectrographs and new imaging data from Keck/NIRC2. Our dataset consists of 16 epochs of relative astrometry, $Hipparcos-Gaia$ absolute astrometry spanning a baseline of 25 years, and 532 RVs --- including 328 unpublished RVs acquired between July 2003 and September 2025. We describe these observations in detail in Section \ref{sec:observations} and our analysis and results in Section \ref{sec:analysis}. We discuss the implications of our findings in Section \ref{sec:discussion} and conclude in Section \ref{sec:conclusion}.


\section{Observations}
\label{sec:observations}

\subsection{Radial Velocities}
\label{subsec:rv_observations}

We consider seven precision RV datasets spanning between May 1981 and March 2026.\footnote{Although a few dozen absolute RVs reaching back to 1896 are available in the literature (e.g. \citealt{Griffin2002, Torres07, Mugrauer2014}), we opted to exclude these measurements because they did not contribute significantly to our model.} The first long-term precision RV monitoring of $\gamma$ Cep was conducted at the Canada-France-Hawaii Telescope (CFHT) using a hydrogen fluoride (HF) gas cell placed in the light path to imprint a set of reference absorption lines onto the spectrum \citep{Campbell1979}. \cite{Campbell1988} reported 26 RVs of $\gamma$ Cep collected using this technique between May 1981 and March 1987. \cite{Walker1992} extended and reprocessed this dataset for a total of 69 RVs taken through November 1991 with a median uncertainty of 10.0 m/s.


\cite{Hatzes03} published 131 RVs from the McDonald Observatory Planet Search (MOPS) program \citep{CochranHatzes2004} taken with three completely different configurations of the McDonald Observatory 2.7m Tull coud\'{e} Spectrograph between July 1988 and December 2002. MOPS Phases I and II used the McDonald 2.7m coud\'{e} ``six-foot camera" (TS12) with a $800\times800$ pixel Texas Instruments CCD detector. The 43 MOPS Phase I $\gamma$ Cep RVs, obtained between July 1988 and November 1994, used telluric O$_2$ lines as the velocity metric \citep{GriffinGriffin1973}. MOPS Phase II, which contributes 49 $\gamma$ Cep RVs between November 1990 and January 1998, also used the same spectrograph configuration but with a temperature stabilized I$_2$ gas cell as the velocity metric \citep{KochWoehl1984}. MOPS Phase III used a totally different 2.7m Tull coud\'{e} Spectrograph, which was a cross-dispersed white-pupil echelle spectrograph with a Tektronix $2048\times2048$ detector \citep{Tull1995}, with the same I$_2$ gas absorption cell. Observations of $\gamma$ Cephei using the ``TS23" spectrograph started in July 1998 and are ongoing as of 2026. 




To these published RVs we contribute four new datasets. The first is a re-reduction of all McDonald Phase III RVs to date using the procedure described in \cite{Endl2000}. These include the data published in \cite{Hatzes03}, plus 210 RVs collected between 2003 and 2025, for a total of 253 observations with a median uncertainty of $\sigma_{RV}=5.2$ m/s.\footnote{\cite{Hatzes03} presented 39 Phase III RVs between July 1998 and December 2002, whereas we present 43 RVs in that same period. The discrepancy arises from differences in the binning procedures during RV reduction.}

The second dataset comprises 40 RVs with uniform uncertainties of 40 m/s from the TRES spectrograph \citep{SzentgyorgyiFuresz2007} at the F. L. Whipple Observatory on Mount Hopkins, AZ between July 2009 and January 2018. The instrument delivers a resolving power of $\lambda/\Delta\lambda \approx$ 44,000 in 51 orders, covering the wavelength range 3800--9100~\AA. The signal-to-noise ratios of the spectra range from about 150 to 600 per resolution element of 6.8 km/s. The reductions were performed with a dedicated pipeline \citep[see][]{Buchhave2012}, and the velocity zero-point was monitored with observations of IAU standards each run, and transferred to the IAU system using observations of minor planets. Radial velocities were determined by cross-correlation using a synthetic template selected from a large pre-computed library of synthetic spectra based on model atmospheres by R.\ L.\ Kurucz \citep[see][]{Nordstrom1994, Latham2002}. The best results were obtained with a template with $T_{\rm eff} = 5000$~K and $\log g = 3.5$.

The third RV dataset includes five precision RVs ($\sigma_{RV}\sim1.0$ m/s) from the HIRES spectrometer \citep{Vogt1994} at Keck Observatory, taken between July 2010 and July 2015. Our team obtained and reduced the HIRES RVs according to the standard procedure of the California Planet Search (CPS; \citealt{Howard2010}). Briefly, we placed a warmed I$_2$ cell in front of the spectrometer entrance slit for each observation to imprint a set of absorption lines with known wavelengths directly onto the observed spectrum as a reference. We then forward modeled the observed star-iodine spectrum by multiplying a high signal-to-noise, iodine-free ``template" of the target star by a laboratory iodine spectrum and convolving this product with the line spread function (LSF) of the spectrometer/telescope system. We measured the LSF each night by observing the featureless spectrum of a rapidly rotating B star through the I$_2$ cell. 

The fourth and final time series consists of 73 RVs from the Levy spectrometer \citep{Vogt2014} mounted on the the Automated Planet Finder telescope at Lick Observatory. These measurements span from October 2013 to March 2026 and have uncertainties of $\sigma_{RV}=1.9$ m/s. The procedure used to reduce APF RVs, described in \cite{Fulton2015}, is analogous to that used for HIRES. The new RVs presented in this work can be found in Table \ref{tab:radial_velocities}.

\begin{deluxetable}{cccc}
\tabletypesize{\footnotesize}
\tablecolumns{8}
\tablewidth{0pt}
\tablecaption{Radial Velocities}

\label{tab:radial_velocities}

\tablehead{
    \colhead{BJD$_{\text{TDB}}$-2400000.0} &
    \colhead{RV (m/s)} &
    \colhead{RV Error (m/s)} &
    \colhead{Instrument}
    }
\startdata
51010.3967 & 11278.3 &    4.5 &  McDonald III \\
51010.4015 & 11283.1 &    4.5 &  McDonald III \\
51065.3476 & 11323.2 &    4.9 &  McDonald III \\
55015.4818 &  -43660 &     40 &          TRES \\
55018.4475 &  -43700 &     40 &          TRES \\
55167.1551 &  -43630 &     40 &          TRES \\
55396.6345 &  -226.5 &    0.7 &         HIRES \\
55544.1796 &  -165.9 &    1.1 &         HIRES \\
56888.5844 &    97.1 &    1.0 &         HIRES \\
56589.2867 &    85.3 &    1.9 &           APF \\
56597.2340 &    67.9 &    1.9 &           APF \\
56606.2020 &    82.9 &    2.0 &           APF \\
\enddata
\tablecomments{Sample RVs from each of the four new data sets presented in this work. All RVs are available in machine-readable format in the online version of this article.}
\end{deluxetable}

\subsection{Relative Astrometry}
\label{subsec:archival_relAst}

We fit 16 epochs of high-resolution direct imaging of $\gamma$ Cep A and B, including eight literature measurements and eight epochs presented here for the first time. The earliest published relative astrometry of $\gamma$ Cep B was presented by \cite{Neuhauser07}, who collected three observations in the $K$-band between July and September 2006 with the Coronagraphic Imager for Adaptive Optics (CIAO; \citealt{Tamura2000}) on the Subaru telescope, and the $\Omega$ Cass IR imager \citep{Lenzen1998} on the 3.5-meter telescope at Calar Alto Observatory. These observations provided the first measurement of $\gamma$ Cep B's apparent magnitude: $K=7.3 \pm 0.2$.

More recently, \cite{Mugrauer22} published five epochs of imaging in the SDSS $i'$-band (centered on 7700 \AA; \citealt{Fukugita1996}) between August 2014 and September 2020. They observed $\gamma$ Cep A and B with the AstraLux lucky imager at Calar Alto Observatory.

\begin{figure*}[t]
    \centering
    \includegraphics[width=\linewidth]{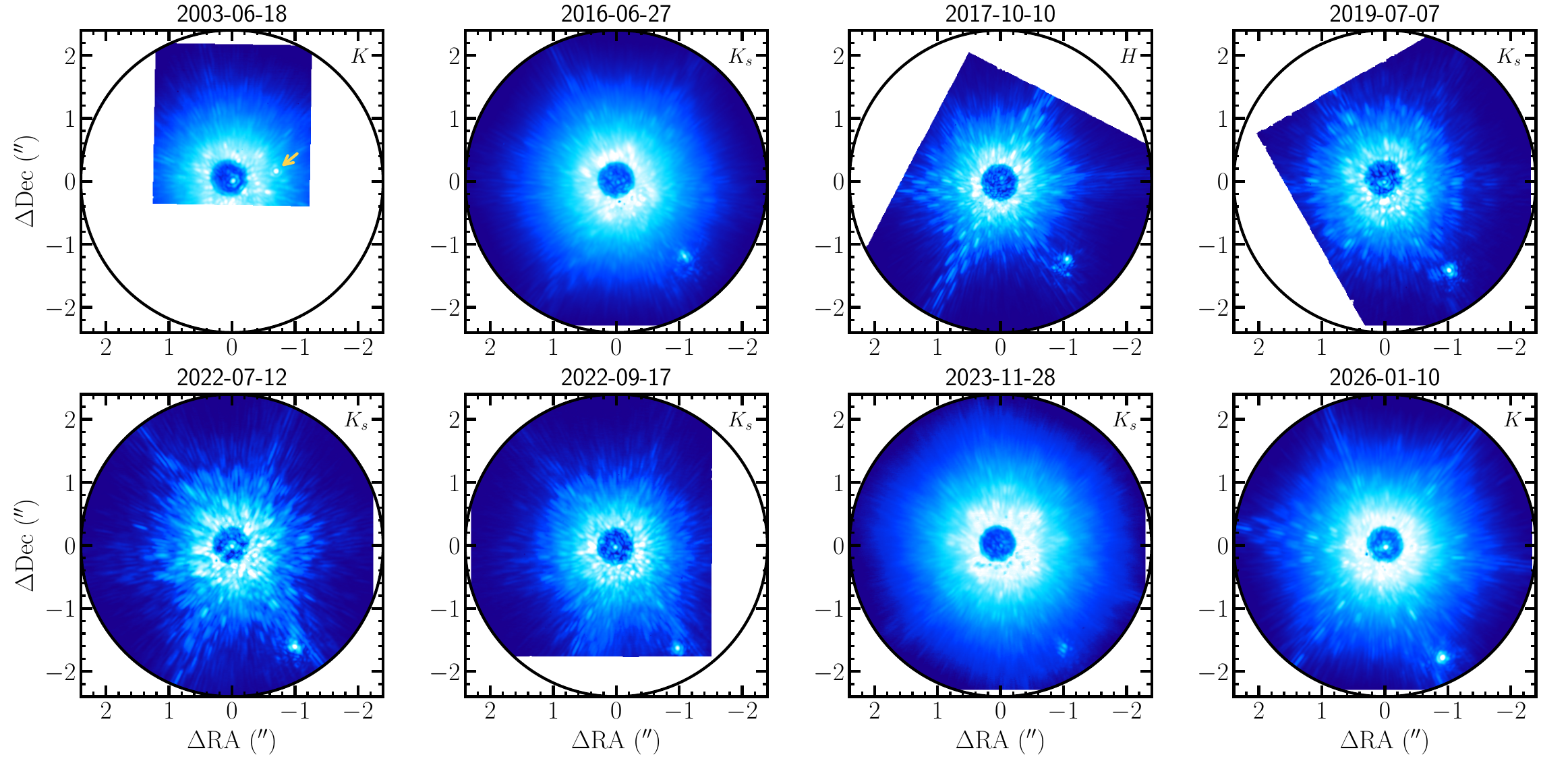}
    \caption{Keck/NIRC2 high-contrast imaging of $\gamma$ Cep A and B between 2003 and 2026. Each image is oriented so that north is up and east is to the left. For the 2003 epoch, the position of $\gamma$ Cep B is indicated with an arrow.}
    \label{fig:nirc2_imaging}
\end{figure*}

We include in our fit eight new adaptive optics (AO) imaging epochs collected using the NIRC2 camera at the 10-m Keck II telescope. We monitored $\gamma$ Cep AB with Keck/NIRC2 between 2016 and 2026 using the narrow camera mode behind the natural guide star adaptive optics system \citep{Wizinowich2000}. An early epoch from 2003 (PI: M.~Liu) extends the astrometric baseline to 23 yr, spanning roughly one third of the binary orbital period. Observations were obtained over eight epochs in $H$, $K$, or $K_S$ bands with an average cadence of approximately 1--2 yr for the past decade. Standard image processing is applied to each science frame, including bad pixel correction, flat fielding, and correction for optical distortions. For observations obtained prior to the Keck AO realignment in 2015 April, we adopt the geometric distortion solution of \cite{Yelda2010}, while epochs obtained after the upgrade use the updated solution from \cite{Service2016}. Relative astrometry is measured from the individual reduced frames by fitting the centroids of both stellar components, and the final separation and position angle at each epoch are computed from the mean of the measurements. Uncertainties are estimated from the standard deviation of the individual frames and include contributions from centroiding precision, plate scale calibration, north alignment uncertainties, and residual distortion errors following the prescription adopted in \cite{Bowler2018}. The resulting measurements reveal substantial orbital motion over the 23 yr baseline, with the projected separation increasing from 689 mas in 2003 to 1979 mas in 2026 while the position angle decreased from $282\fdg7$ to $207\fdg1$. Details of the new NIRC2 observations can be found in Table \ref{tab:nirc2}, with individual epochs displayed in Figure~\ref{fig:nirc2_imaging}. Summary information for all 16 epochs used in our analysis appears in Table \ref{tab:relative_astrometry}.

\begin{deluxetable*}{lccccc}
\renewcommand\arraystretch{0.9}
\tabletypesize{\small}
\setlength{ \tabcolsep } {.1cm} 
\tablewidth{0pt}
\tablecolumns{6}
\tablecaption{Keck/NIRC2 Adaptive Optics Imaging of $\gamma$ Cep AB \label{tab:nirc2}}
\tablehead{
       \colhead{UT Date} &  \colhead{Epoch}  & \colhead{$N$$\times$Coadds$\times$$t_\mathrm{exp}$}  & \colhead{Filter\tablenotemark{a}}  & \colhead{Sep.}  & \colhead{P.A.}    \\
      &  \colhead{(UT)}       & \colhead{(s)}                  & \colhead{}                      &  \colhead{($''$)}       & \colhead{ ($^{\circ}$)}           
        }   
\startdata
2003 Jun 18  & 2003.4620  &  10 $\times$ 100 $\times$ 0.035  &  $K$+cor600  &  0.689 $\pm$ 0.003  &  282.7 $\pm$ 0.1      \\
2016 Jun 27  & 2016.4879  &  5 $\times$ 50 $\times$ 0.2  &  $K_s$+cor600  &  1.585 $\pm$ 0.004  &  222.0 $\pm$ 0.2      \\
2017 Oct 10  & 2017.7738  &  20 $\times$ 50 $\times$ 0.53  &  $H$+cor600  &  1.654 $\pm$ 0.005  &  218.8 $\pm$ 0.1      \\
2019 Jul 07  & 2019.5138  &  20 $\times$ 100 $\times$ 0.53  &  $K_s$+cor600  &  1.719 $\pm$ 0.010  &  216.8 $\pm$ 0.2      \\
2022 Jul 12  & 2022.5273  &  10 $\times$ 10 $\times$ 0.181  &  $K_s$+cor600  &  1.889 $\pm$ 0.004  &  211.9 $\pm$ 0.1      \\
2022 Sep 17  & 2022.7104  &  10 $\times$ 10 $\times$ 0.1  &  $K_s$+cor600  &  1.885 $\pm$ 0.004  &  211.5 $\pm$ 0.1      \\
2023 Nov 28  & 2023.9073  &  4 $\times$ 1 $\times$ 3  &  $K_s$+cor600  &  1.886 $\pm$ 0.021  &  209.6 $\pm$ 0.5      \\
2026 Jan 10  & 2026.0251  &  5 $\times$ 50 $\times$ 0.17  &  $K$+cor600  &  1.979 $\pm$ 0.007  &  207.1 $\pm$ 0.2      \\
\enddata
\tablenotetext{a}{``cor600'' refers to the 600 mas-diameter focal plane coronagraph.}
\end{deluxetable*}
\begin{deluxetable}{ccccc}
\tabletypesize{\footnotesize}
\tablecolumns{8}
\tablewidth{0pt}
\tablecaption{Relative Astrometry of $\gamma$ Cep B}

\label{tab:relative_astrometry}

\tablehead{
    \colhead{Date} &
    \colhead{$\rho$ (arcsec)} &
    \colhead{PA ($^{\circ}$)} &
    \colhead{Instrument} &
    \colhead{Source}
    }
\startdata
2003.462  & 0.689 $\pm$  0.003  & 282.7  $\pm$  0.1  &   NIRC2  & 2 \\
2006.536 & 0.870 $\pm$  0.005 & 256.91 $\pm$  0.27 & CIAO & 1 \\
2006.693 & 0.891 $\pm$  0.006 & 256.16 $\pm$  0.35 & $\Omega$ Cass & 1 \\
2006.696 & 0.887 $\pm$  0.005 & 256.39 $\pm$  0.34 & $\Omega$ Cass & 1 \\
2014.635 & 1.441 $\pm$  0.011 & 226.05 $\pm$  0.41 & AstraLux & 3 \\
2015.652 & 1.517 $\pm$  0.011 & 223.89 $\pm$  0.41 & AstraLux & 3 \\
2016.487  & 1.585 $\pm$  0.004  & 222.0  $\pm$  0.2  &   NIRC2  & 2 \\
2016.821 & 1.603 $\pm$  0.014 & 220.32 $\pm$  0.49 & AstraLux & 3 \\
2017.773  & 1.654 $\pm$  0.005  & 218.8  $\pm$  0.1  &   NIRC2  & 2 \\
2017.827 & 1.652 $\pm$  0.008 & 219.70 $\pm$  0.29 & AstraLux & 3 \\
2019.513  & 1.719 $\pm$  0.010  & 216.8  $\pm$  0.2  &   NIRC2  & 2 \\
2020.679 & 1.787 $\pm$  0.007 & 214.60 $\pm$  0.23 & AstraLux & 3 \\
2022.527  & 1.889 $\pm$  0.004  & 211.9  $\pm$  0.1  &   NIRC2  & 2 \\
2022.710  & 1.885 $\pm$  0.004  & 211.5  $\pm$  0.1  &   NIRC2  & 2 \\
2023.907  & 1.886 $\pm$  0.021  & 209.6  $\pm$  0.5  &   NIRC2  & 2 \\
2026.025  & 1.979 $\pm$  0.007  & 207.1  $\pm$  0.2  &   NIRC2  & 2 \\
\enddata
\tablerefs{(1) \cite{Neuhauser07}, (2) This work, (3) \cite{Mugrauer22}}
\end{deluxetable}

\subsection{Absolute Astrometry}
\label{subsec:absolute_astrometry}

We used \texttt{orvara}'s built-in compatibility with the $Hipparcos$-$Gaia$ Catalog of Accelerations (HGCA; \citealt{Brandt2021a}) to incorporate absolute astrometric constraints. The HGCA provides three values of astrometric proper motion: one measured at the $Hipparcos$ mission epoch ($\sim$1991.25), one measured at the $Gaia$ EDR3 epoch ($\sim$2016.0), and one calculated from the change in the stellar position between these two epochs. Because the HGCA placed $Hipparcos$ and $Gaia$ into a common reference frame, differences in these proper motion values, known also as proper motion anomalies or PMa \citep{Kervella2019}, suggest stellar acceleration in the sky plane, from which the presence of a companion might be inferred. 

\begin{figure}[t]
    \centering
    \includegraphics[width=\linewidth]{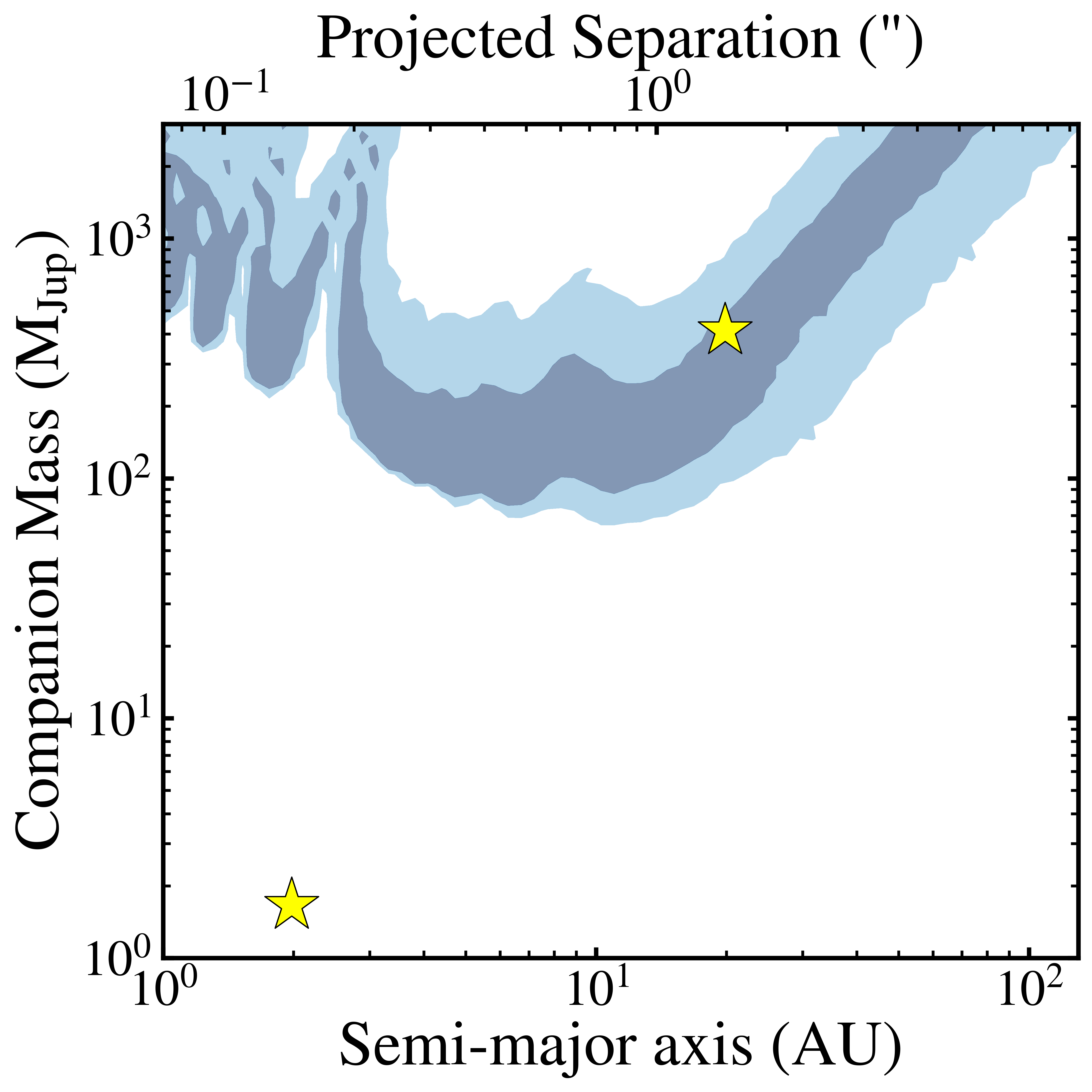}
    \caption{Joint mass-separation posterior for $\gamma$ Cep using absolute astrometry from $Hipparcos$ and $Gaia$. Companions with mass/separation pairs that fall in the blue region would produce astrometric accelerations consistent with that observed in the HGCA. The dark (light) blue contours contain 68\% (95\%) of the posterior mass. The yellow stars mark the parameters of $\gamma$ Cep B (upper right), which accounts for the acceleration, and $\gamma$ Cep Ab (lower left), which contributes about 1\% of the total astrometric signal.}
    \label{fig:astro_ethraid}
\end{figure}

The HGCA reports $Hipparcos$-$Gaia$ average proper motions for $\gamma$ Cep A of $\mu_{\alpha*,HG}=-51.38\pm0.02$ mas/yr and $\mu_{\delta,HG}=168.30\pm0.02$ mas/yr, and $Gaia$ proper motions of $\mu_{\alpha*,G}=-64.9\pm0.2$ mas/yr and $\mu_{\delta,G}=171.2\pm0.2$ mas/yr, with an associated $\chi^2$ of 4771 and SNR of 69.1. From these proper motions we calculated a PMa of $\dmu$ mas/yr, equivalent to an acceleration of $72\pm1$ m/s/yr over the 12.4 years between the $Hipparcos$-$Gaia$ midpoint (2003.6) and the $Gaia$ EDR3 epoch (2016.0). This significant astrometric acceleration is consistent with $\gamma$ Cep A's high renormalized unit weight error (RUWE; \citealt{Lindegren2018, GaiaEDR32021_solution}) of 3.2. RUWE measures the goodness of fit of $Gaia$'s single-star astrometric motion model, with values larger than $\approx1.2$ indicating binarity (e.g., \citealt{Bryson2020, Berger2020}).
We modeled the PMa expected from both $\gamma$ Cep B and $\gamma$ Cep Ab, finding that $\Delta \mu_{\text{B}}\approx13.9$ mas/yr, while $\Delta \mu_{\text{Ab}}\approx0.11$ mas/yr. We therefore conclude that $\gamma$ Cep Ab had a negligible effect on the astrometric signal of $\gamma$ Cep A, as shown in Figure~\ref{fig:astro_ethraid}.

\section{Analysis}
\label{sec:analysis}

\subsection{Joint RV/Astrometry Orbital Fit}
\label{subsec:orbit_fit}

We used \texttt{orvara} \citep{Brandt2021b} to jointly fit the orbits of $\gamma$ Cep B and Ab, which supports joint modeling of RVs, absolute astrometry, and relative astrometry. Our 30-parameter model included the primary mass, system parallax, a jitter term and a zero point offset for each of the seven RV instruments, and seven parameters per companion: mass, semi-major axis, $\sqrt{e}\sin \omega$, $\sqrt{e}\cos \omega$, inclination, longitude of the ascending node, and the mean longitude at a reference epoch $t_{\text{ref}}=2010.0$. We sampled the joint likelihood with the parallel-tempering MCMC sampler \texttt{ptemcee} \citep{Vousden2016}, which runs simultaneous MCMC chains at different temperatures to let walkers more easily navigate multimodal distributions. We ran 100 walkers at 25 different temperatures and for 100,000 steps each. We kept only the lowest-temperature chain from each walker and discarded the first 80,000 steps as burn-in. To ensure satisfactory convergence, we required chains to exceed 50 times the autocorrelation length. We also visually assessed the walkers' exploration of the parameter space, as well as the agreement of our fitted orbital models with our data, using \texttt{orvara}'s diagnostic plots.

We applied a Gaussian prior on the mass of $\gamma$ Cep A matching the value found by \cite{Knudstrup2023}, $1.27^{+0.05}_{-0.07}~\Msun$.\footnote{We tested the effect of this prior by repeating our fit with the prior uncertainties inflated by a factor of four. We found that astrophysically interesting parameters such as the primary mass and the masses, separations, and eccentricities of $\gamma$ Cep B and Ab varied within 1$\sigma$, while the uncertainties on those parameters inflated by $50-100\%$.} For fits using absolute astrometry, \texttt{orvara} automatically implements a Gaussian prior using $Gaia$ EDR3 values \citep{GaiaEDR32021_solution}. We set log-uniform priors on companion mass, companion separation, and RV jitter, and a uniform prior on $\sin i$.

\begin{figure*}[t]
    \centering

    \begin{minipage}[b]{0.49\linewidth}
        \centering
        {\includegraphics[width=\linewidth]{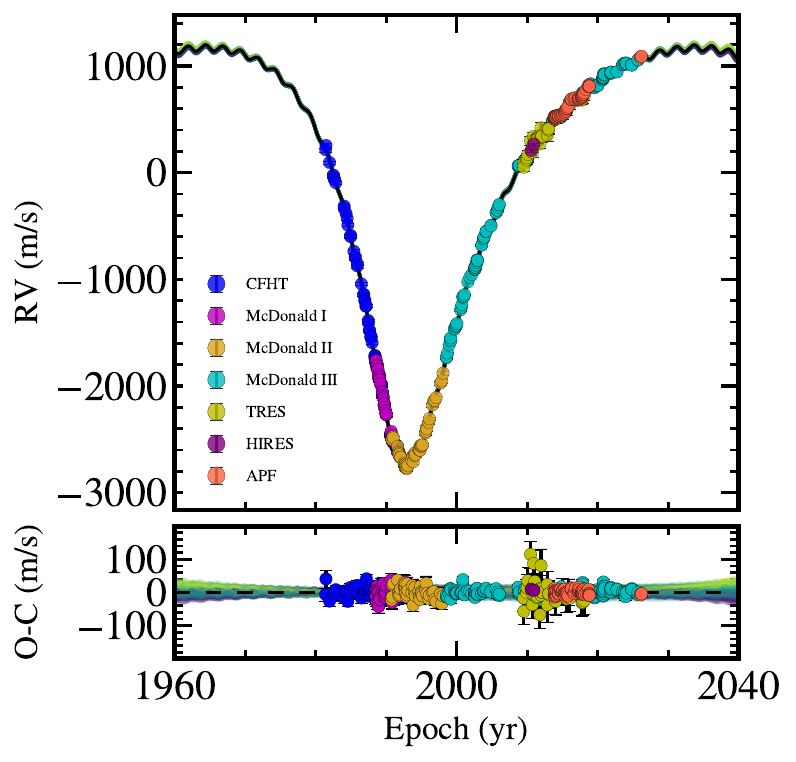}}
    \end{minipage}
    \hfill
    \begin{minipage}[b]{0.49\linewidth}
        \centering
        \includegraphics[width=\linewidth]{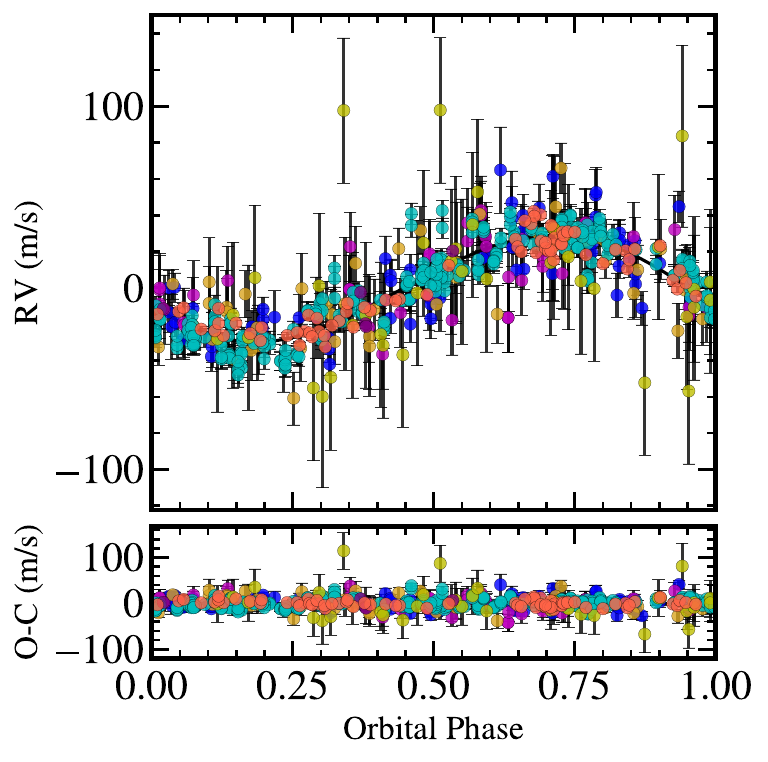}
    \end{minipage}

    \caption{Radial velocity measurements of the $\gamma$ Cep system between May 1981 and March 2026. \textbf{Left:} Full RV model including $\gamma$ Cep B and $\gamma$ Cep Ab, with symbols colored according to the instrument with which they were measured. Orbital posterior draws are colored according to the eccentricity of $\gamma$ Cep B, with lighter colors indicating higher eccentricity. The thick black line shows the maximum likelihood model fit, and residuals to this fit are shown in the bottom panel. \textbf{Right:} Residuals after subtracting the signature of $\gamma$ Cep B. RV measurements are colored as in the right figure and phase-folded to the period of $\gamma$ Cep Ab. Residuals are shown in the bottom panel.}
    \label{fig:rv_orbit}
\end{figure*}

\begin{figure*}[htbp]
\includegraphics[width=\linewidth]{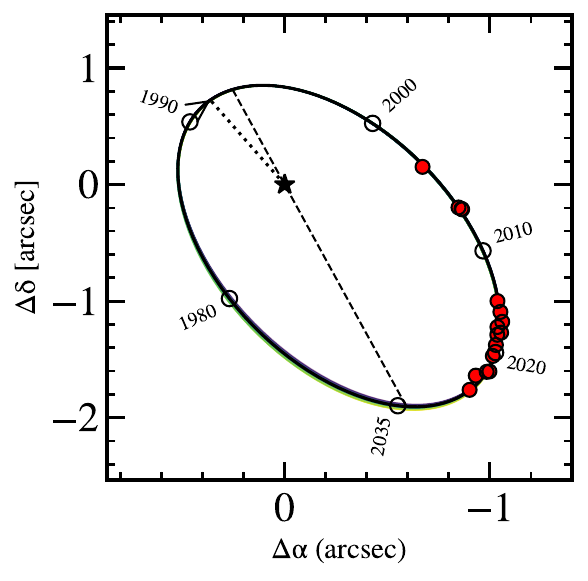}
    \caption{Relative astrometry and orbit models of $\gamma$ Cep B's orbit around $\gamma$ Cep A (black star). Red circles depict the position of $\gamma$ Cep B relative to $\gamma$ Cep A between June 2003 and January 2026. Unfilled circles show predicted positions at the epochs indicated. Colored lines show posterior draws of $\gamma$ Cep B's orbit, with hue corresponding to $\gamma$ Cep B's mass. The dotted line connects the host star to $\gamma$ Cep B's periastron passage, and the dashed line indicates the line of nodes.}
    \label{fig:astro_orbit}
\end{figure*}

\begin{figure*}[htbp]
    \centering
    \begin{subfigure}
        \centering
        \includegraphics[width=0.49\linewidth]{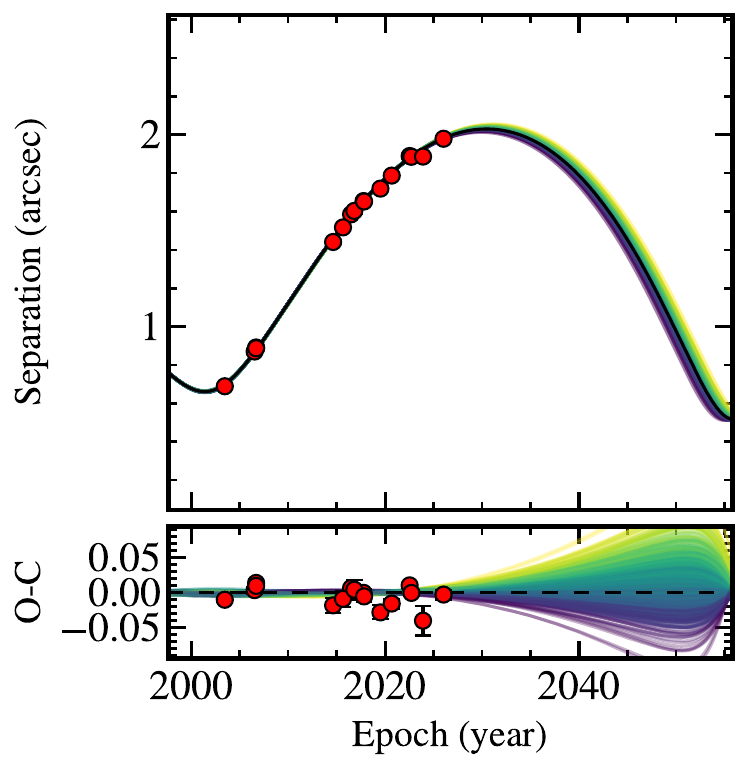}
    \end{subfigure}
    \begin{subfigure}
        \centering
        \includegraphics[width=0.47\linewidth]{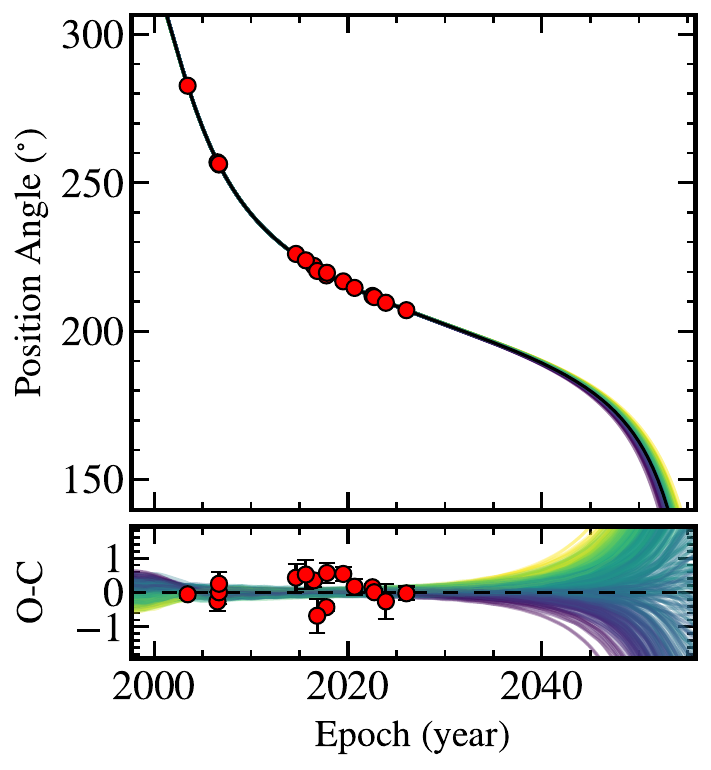}
    \end{subfigure}

    \vspace{1em} 
    \begin{subfigure}
        \centering
        \includegraphics[width=0.48\linewidth]{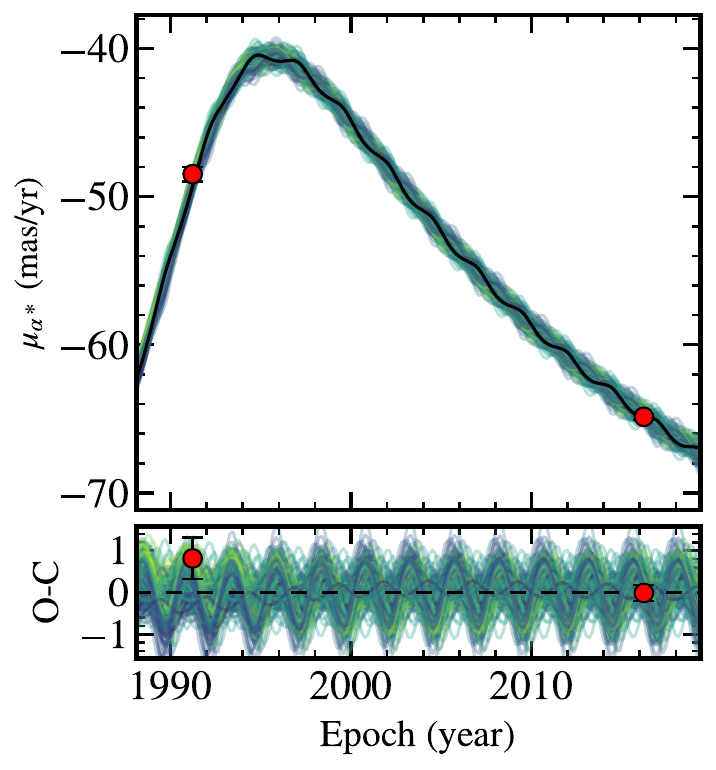}
    \end{subfigure}
    \begin{subfigure}
        \centering
        \includegraphics[width=0.48\linewidth]{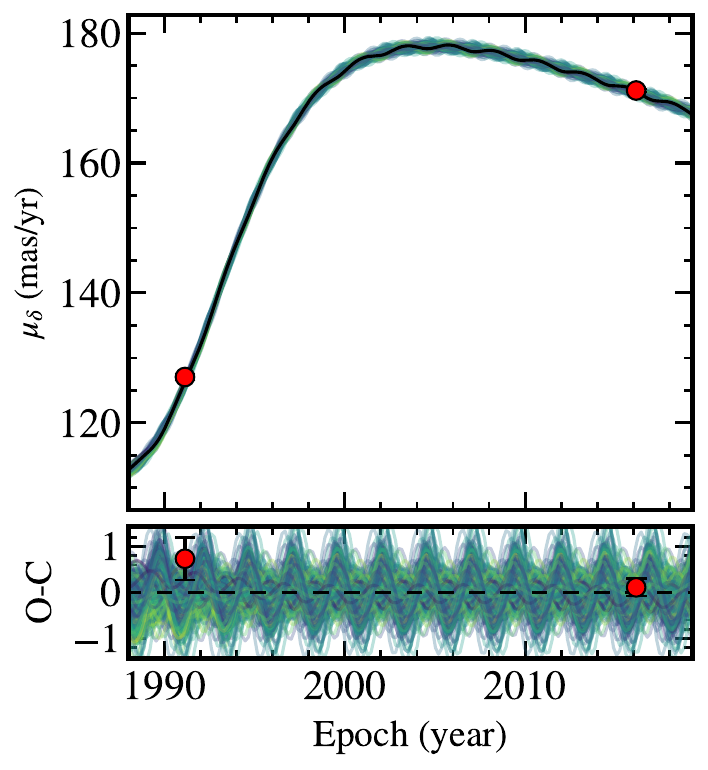}
    \end{subfigure}

    \caption{Observed (red circles) and modeled (colored lines) relative and absolute astrometry of the $\gamma$ Cep system. The top row shows the relative position of $\gamma$ Cep B with respect to $\gamma$ Cep A --- also shown in Figure \ref{fig:astro_orbit} --- as a function of time. The bottom row shows absolute proper motion measurements of $\gamma$ Cep A from the $Hipparcos$ and $Gaia$ missions as red circles. The mean proper motion between the two missions also constrains the orbit, but is not shown in the plots. In all subplots, colored orbits show posterior draws from our fit, the black line shows the maximum-likelihood orbital fit, and the lower panel shows residuals of both the data and the posterior draws to the maximum-likelihood fit. Additional AO imaging observations in 2040-2050 would help to further constrain the orbit.}
    \label{fig:four_panel_astrometry}
\end{figure*}

The parameters we derived for the $\gamma$ Cep system are shown in Table \ref{tab:system_params}. Model fits to the radial velocities, relative astrometry, and absolute astrometry are shown in Figures \ref{fig:rv_orbit}, \ref{fig:astro_orbit}, and \ref{fig:four_panel_astrometry}, respectively. The orbital parameters we derived for both $\gamma$ Cep B and Ab are in general agreement with those reported in the literature (see Tables \ref{tab:parameter_comparison} and \ref{tab:inc_Om_om_comparison}), with a few notable exceptions. 

First, our derived period and mass for $\gamma$ Cep B are within about $2\sigma$ of those found by \cite{Torres07}, \cite{Neuhauser07}, and \cite{Mugrauer22}. This marginal agreement is not unexpected, given our incorporation of data from these studies alongside new RVs and imaging in our analysis. By contrast, we find $3-5\sigma$ differences between our values of $M_B$ ($\MBsolar~\Msun$), $M_{Ab} \sin i$ ($\MAbsini~\Mjup$), and $P_{Ab}$ ($\PAb$ years) and those found by \cite{Knudstrup2023} ($M_B = 0.328^{+0.009}_{-0.012}~\Msun$, $M_{Ab} \sin i = 1.41^{+0.08}_{-0.08}~\Mjup$, $P_{Ab} = 2.500^{+0.008}_{-0.008}$ years). The most significant discrepancy was between our derived values of the binary period $P_B$, which we measured to be $\PB$ years, versus \cite{Knudstrup2023}'s value of $58.0^{+0.1}_{-0.2}$ years, a difference of $\sim30\sigma$. This disagreement likely arises from their use of a completely distinct set of high-cadence RVs from the SONG network \citep{Grundahl2008}, our incorporation of multiple decades of adaptive optics imaging observations, and differences in priors and modeling approach.

Second, the planetary inclination value inferred from our joint \texttt{orvara} fit ($i=\iAb^{\circ}$) is dominated by our prior, a sine distribution symmetric about $90^{\circ}$, indicating no preference in the data for an inclined orbit. We tested this outcome by performing a second fit in which $i_{Ab}$ was sampled uniformly between the bounds found by \cite{ReffertQuirrenbach2011}: $2.8^{\circ}$--$20.8^{\circ}$ and $166.6^{\circ}$--$174.8^{\circ}$. We found in both cases a consistent preference for walkers to hug the bound of the range closest to $90^{\circ}$, indicating a tension between our likelihood and the \cite{ReffertQuirrenbach2011} prior.



Finally, we note a difference in convention between our fitted values and those reported in the literature for the argument of periastron $\omega$ and the longitude of the ascending node $\Omega$ of the binary and planetary orbits. \texttt{orvara} defines a negative RV value as corresponding to an object receding from the observer \citep{Householder2022}, whereas some past works have used the opposite convention. This change affects the value of $\Omega$ and, by extension $\omega$. For consistency, we adopt the latter convention and use it in all reported values of $\omega$ and $\Omega$ throughout this work. Our correction for this difference is two-fold. First, we apply a $180^{\circ}$ offset to our derived $\Omega$ value for $\gamma$ Cep B or Ab. Second, because $\omega$ is defined in reference to the ascending node, we must likewise offset $\omega$ by $180^{\circ}$ to maintain the same angular separation between them.



\texttt{orvara} defines the ascending node as the point where the orbiting body passes through the sky plane while moving toward the observer. By contrast, most previous works defined $\Omega$ as the point where the orbiting body passes through the sky plane while moving \textit{away} from the observer. For consistency, we adopt the latter convention and use it in all reported values of $\Omega$ throughout this work. To correct for this difference, we apply a $180^{\circ}$ offset to our derived $\Omega_B$ value. We do the same for $\Omega_{Ab}$.

Measurements of $\gamma$ Cep Ab's minimum mass, $M_{Ab}\sin i$, by past studies have generally fallen between $
\approx 1.3-2.0~\Mjup$. Our derived value of $\MAbsini \, \Mjup$ falls squarely within this interval. Notably, however, the two studies that derived the closest host star mass values to our own value of $\MA~\Msun$ --- \citet{Torres07} ($1.18 \pm 0.11 \, M_{\odot}$) and \citet{Knudstrup2023} ($1.27^{+0.05}_{-0.07} \, M_{\odot}$) --- found $M_{Ab} \sin i$ values in tension with ours at the $\sim2\sigma$ level ($1.43\pm0.13~\Mjup$ and $1.41\pm0.08~\Mjup$, respectively). We also found a lower eccentricity for $\gamma$ Cep Ab, $e_{Ab}=\eAb$, than all past work other than \citet{Endl2011} ($0.049\pm0.034$), though large uncertainties render most measurements consistent at the $1\sigma$ level.

\begin{table*}[t]
\centering
\caption{System Parameters}
\label{tab:system_params}
\begin{tabular}{lccccc}
\hline
\hline
\multicolumn{6}{c}{\textbf{Stellar Parameters}} \\
\hline
\textbf{Parameter} & \textbf{Value} & \textbf{Source} &  &  &  \\
\hline
\textbf{General} &  &  &  &  &  \\
Other Names &  HD 222404, HIP 116727, HR 8974 & ---\\
RA (2016.0) & 23:39:20.59 & \citet{GaiaEDR32021_solution}\\
Dec (2016.0) &  +77:37:59.2 & \citet{GaiaEDR32021_solution} \\
V mag &  3.21 & \citet{Torres07} \\
G mag &  2.94 & \citet{GaiaEDR32021_solution} \\
Mass (\Msun) & \MA & \citet{Knudstrup2023} and this work\\
Metallicity ([Fe/H]) & \FeHA & \citet{Knudstrup2023}\\
\hline
\textbf{Astrometry} \\
Parallax (mas) & 72.5 $\pm$ 0.1 & \citet{GaiaEDR32021_solution} \\
Proper Motion in RA (mas/yr) & -64.9 $\pm$ 0.1 & \citet{GaiaEDR32021_solution} \\
Proper Motion in Dec (mas/yr) & 171.2 $\pm$ 0.1 & \citet{GaiaEDR32021_solution} \\
Radial Velocity (km/s) & -43.668 $\pm$ 0.029 & \citet{Bailer-Jones2018} \\
\hline
\multicolumn{6}{c}{\textbf{Companion Parameters}}     \\
\hline
\textbf{Parameter} & \textbf{B} & \textbf{Ab} \\
\hline
Semi-major axis (AU) & \aB & \aAb\\
Orbital Period (years) & \PB & \PAb\\
Eccentricity & \eB & \eAb\\
$\omega$ (degrees) & \oB & \oAb \\
Inclination (degrees) & \iB & \iAb\\
$\Omega$ (degrees) & \OB & \OAb\\
$t_{\text{peri}}$ (days) & \tpB & \tpAb\\
M ($\Mjup$) & \MBjup & ---\\
M$\sin i$ ($\Mjup$) & --- & \MAbsini\\
\hline
\hline
\multicolumn{6}{c}{\textbf{Other Model Parameters}}  \\
\hline
\textbf{Parameter} & \textbf{Value} & &  \\
\hline
$\sigma_{\text{CFHT}}$ (m/s) & $11_{-2}^{+2}$ & ---\\
$\sigma_{\text{McDonald I}}$ (m/s) & $0.010_{-0.009}^{+1.000}$ & ---\\
$\sigma_{\text{McDonald II}}$ (m/s) & $8_{-8}^{+4}$ & ---\\
$\sigma_{\text{McDonald III}}$ (m/s) & $7.4_{-0.5}^{+0.6}$ & ---\\
$\sigma_{\text{TRES}}$ (m/s) & $0.009_{-0.009}^{+1.200}$ & ---\\
$\sigma_{\text{HIRES}}$ (m/s) & $10_{-3}^{+6}$ & ---\\
$\sigma_{\text{APF}}$ (m/s) & $6.4_{-0.6}^{+0.7}$ & ---\\
$\gamma_{\text{CFHT}}$ (m/s) & $\equiv 0$ & ---\\
$\gamma_{\text{McDonald I}}$ (m/s) & $-742 \pm 9$ & ---\\
$\gamma_{\text{McDonald II}}$ (m/s) & $-937 \pm 8$ & ---\\
$\gamma_{\text{McDonald III}}$ (m/s) & $-11419 \pm 8$ & ---\\
$\gamma_{\text{TRES}}$ (m/s) & $45345 \pm 8$ & ---\\
$\gamma_{\text{HIRES}}$ (m/s) & $2028 \pm 8$ & ---\\
$\gamma_{\text{APF}}$ (m/s) & $2033 \pm 8$ & ---\\
\hline
\hline   
\end{tabular}
\end{table*}

\begin{deluxetable*}{lcccccccc}
\tablecaption{Orbital Parameters of $\gamma$~Cep from Selected Studies in the Literature}
\label{tab:parameter_comparison}
\tablehead{
\colhead{Source} &
\colhead{$M_A$} &
\colhead{$M_B$} &
\colhead{$P_B$} &
\colhead{$e_B$} &
\colhead{$M_{Ab} \sin i$} &
\colhead{$M_{Ab}$} &
\colhead{$P_{Ab}$} &
\colhead{$e_{Ab}$} \\
\colhead{} &
\colhead{($\Msun$)} &
\colhead{($\Msun$)} &
\colhead{(yr)} &
\colhead{} &
\colhead{($\Mjup$)} &
\colhead{($\Mjup$)} &
\colhead{(yr)} &
\colhead{}
}
\startdata
\cite{Campbell1988} & --- & --- & --- & --- & --- & --- & 2.7 & --- \\
\cite{Walker1992} & --- & --- & $\geq$30 & $0.21 \pm 0.08$ & 1.4 & --- & 2.5 & --- \\
\cite{Hatzes03}\tablenotemark{a} & $1.59 \pm 0.12$ & 0.34 & $57 \pm 4$ & $0.36 \pm 0.02$ & $1.7 \pm 0.4$ & --- & $2.48 \pm 0.01$ & $0.12 \pm 0.05$ \\
\cite{Torres07} & $1.18 \pm 0.11$ & $0.36 \pm 0.02$ & $66.8 \pm 1.4$ & $0.409 \pm 0.007$ & $1.43 \pm 0.13$ & $\leq$16.9 & $2.47 \pm 0.01$ & $0.11 \pm 0.06$ \\
\cite{Neuhauser07} & $1.40 \pm 0.12$ & $0.409 \pm 0.018$ & $67.5 \pm 1.4$ & $0.4112 \pm 0.0063$ & $1.60 \pm 0.13$ & --- & $2.47 \pm 0.01$ & $0.12 \pm 0.06$ \\
\cite{Endl2011}\tablenotemark{b} & --- & --- & --- & --- & $1.85 \pm 0.16$ & --- & $2.473 \pm 0.004$ & $0.049 \pm 0.034$ \\
\cite{Mugrauer22} & $1.29 \pm 0.08$ & $0.39 \pm 0.03$ & $66.84 \pm 1.32$ & $0.4144 \pm 0.0066$ & --- & --- & --- & --- \\
\cite{Knudstrup2023} & $1.27^{+0.05}_{-0.07}$ & $0.328^{+0.009}_{-0.012}$ & $58.0^{+0.1}_{-0.2}$ & $0.333 \pm 0.002$ & $1.41 \pm 0.08$ & $6.6^{+2.3}_{-2.8}$ & $2.500 \pm 0.008$ & $0.15^{+0.07}_{-0.05}$ \\
This work\tablenotemark{c} & \MA & \MBsolar & \PB & \eB & \MAbsini & --- & \PAb & \eAb \\
\enddata
\tablenotetext{a}{This study quoted the $M_A$ value from \cite{Fuhrmann2004}. Also, their $M_B$ value is actually minimum mass $M_B \sin i$, which we calculated from the reported mass function value, $f(M)=0.0106 \, \Msun$.}
\tablenotetext{b}{This study adopted $M_A$ and $P_B$ values from \cite{Neuhauser07}.}
\tablenotetext{c}{Relaxing the prior on $M_A$ to $1.27^{+0.25}_{-0.25}$, an increase in $\sigma_{M_A}$ of $\sim400\%$, had a modest effect on our derived uncertainties. For example, $\sigma_{M_B}$ increased by 25\%, $\sigma_{P_B}$ by 75\%, and $\sigma_{e_B}$ by 100\%, while all uncertainties on $\gamma$ Cep Ab's properties were negligibly affected.}
\tablenotetext{}{%
\textbf{Note.} This table is meant to be illustrative of past work, not a comprehensive account of all studies of the $\gamma$~Cep system.
}
\end{deluxetable*}

\subsection{Injection/Recovery Tests}
\label{subsec:injection_recovery}

We assessed the possibility of further companions existing in the $\gamma$ Cep system using a version of \texttt{RVSearch} \citep{Rosenthal2021} modified to better detect long-term RV variability \citep{VanZandt2025}. We performed a series of injection/recovery tests on the RV residuals to our fit in Section \ref{subsec:orbit_fit}. In brief, an injection consists of randomly sampling a set of orbital parameters --- in this case the period $P$, periastron time $t_p$, eccentricity $e$, argument of periastron $\omega$, and RV semi-amplitude $K$ --- calculating the associated RV signal at the epoch of each real observation, and adding that value to the residual velocity after subtracting the best fit orbital solution for $\gamma$ Cep B and Ab. The subsequent recovery consists of applying \texttt{RVSearch}'s blind orbit fitting algorithm to the modified residuals, and determining first whether a periodic signal was recovered, and second whether the parameters of the recovered signal match their injected counterparts within tolerance. 

\begin{figure*}[htbp]
\includegraphics[width=\linewidth]{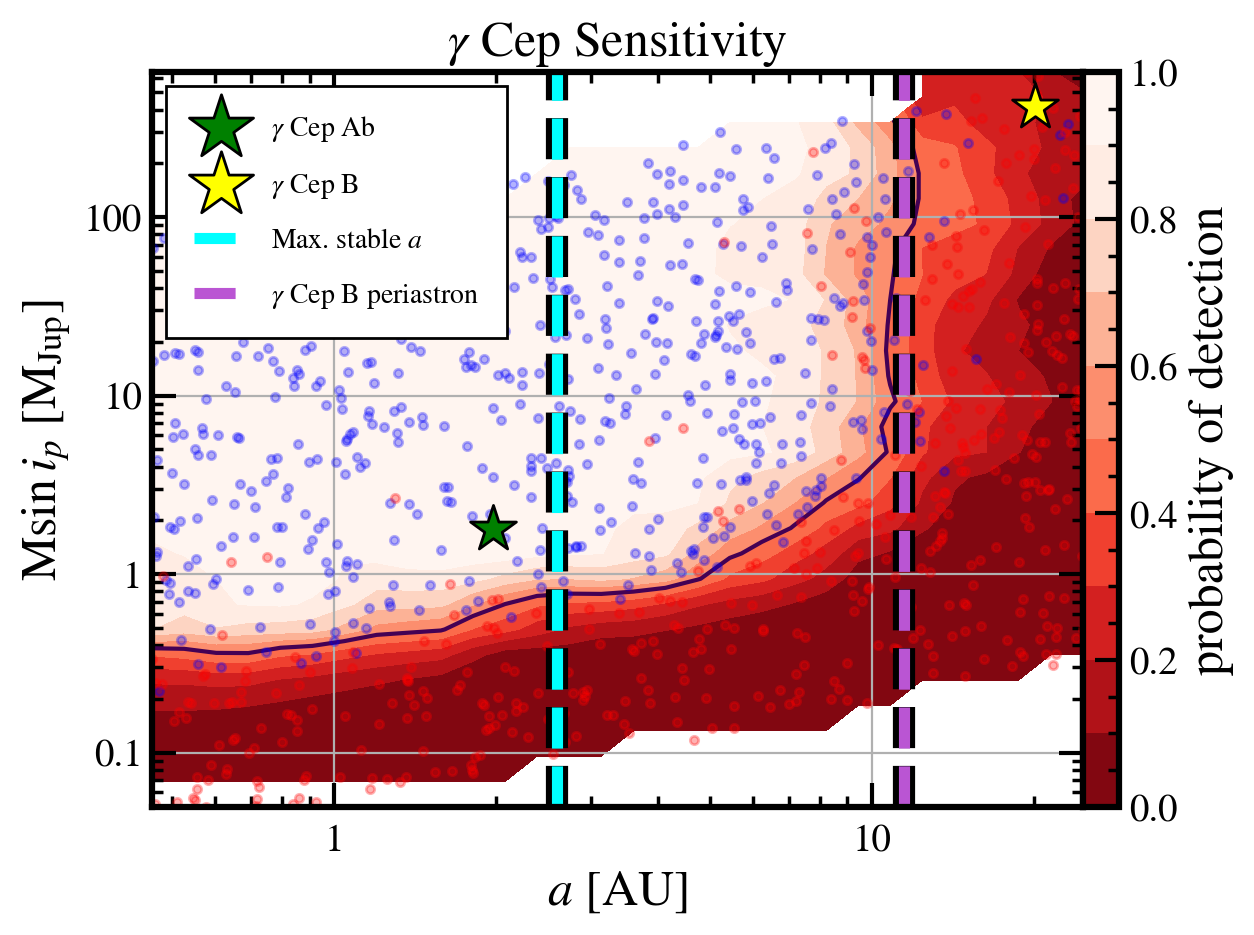}
    \caption{Sensitivity to companions in the $\gamma$ Cep system. Red and blue points represent injected signals that were unsuccessfully and successfully recovered by our blind search algorithm, respectively. The parameter space is shaded according to the local average fraction of successful recoveries, with lighter colors representing higher sensitivity. The black contour shows our 50\% sensitivity limit. We reached 50\% sensitivity to $130 \, \Me$ planets with 100-day periods and 5 $\Mjup$ planets with 28-year periods. The yellow star (upper right) shows the parameters of $\gamma$ Cep B, while the green star (lower left) shows the parameters of $\gamma$ Cep Ab. The purple line marks $\gamma$ Cep B's periastron separation of 11.6 AU, and the cyan line shows the outermost stable orbit of a companion to $\gamma$ Cep A under the assumption of a coplanar system ($\sim2.6$ AU; \citealt{Haghighipour2006}).}
    \label{fig:injection_recovery}
\end{figure*}

We performed 1000 such tests over a range of masses and orbital separations to derive a sensitivity map illustrating which companions would have been detected with RVs, and with what confidence level. We injected orbits with periods between 100--$4\times10^{4}$ days, and with RV semi-amplitudes between 1.5--3000 m/s. This range allowed us to calculate our sensitivity to $M\sin i$ between 13 $\Me$ and 0.8 $\Msun$. We drew eccentricity values from a beta distribution, $\mathcal{B}(0.867, 3.03)$, with parameters fit by \cite{Kipping2013}. We chose a fixed $\Delta$BIC threshold to assess the significance of injections. We determined that a threshold of 30 was appropriate to identify high-significance signals while rejecting marginal ones. In other words, only injected signals satisfying BIC$_{\text{planet}}-$BIC$_{\text{flat}}\geq30$, where the two BIC values are associated with models that respectively include and exclude an additional planet, were considered significant. We achieved 50\% sensitivity to super-Saturns ($\sim130 \, \Me$) at 100 days and super-Jupiters (5 $\Mjup$) at $10^4$ days (10 AU). Our RV sensitivity curve is shown in Figure \ref{fig:injection_recovery}.

\begin{figure*}[htbp]
\includegraphics[width=0.9\linewidth]{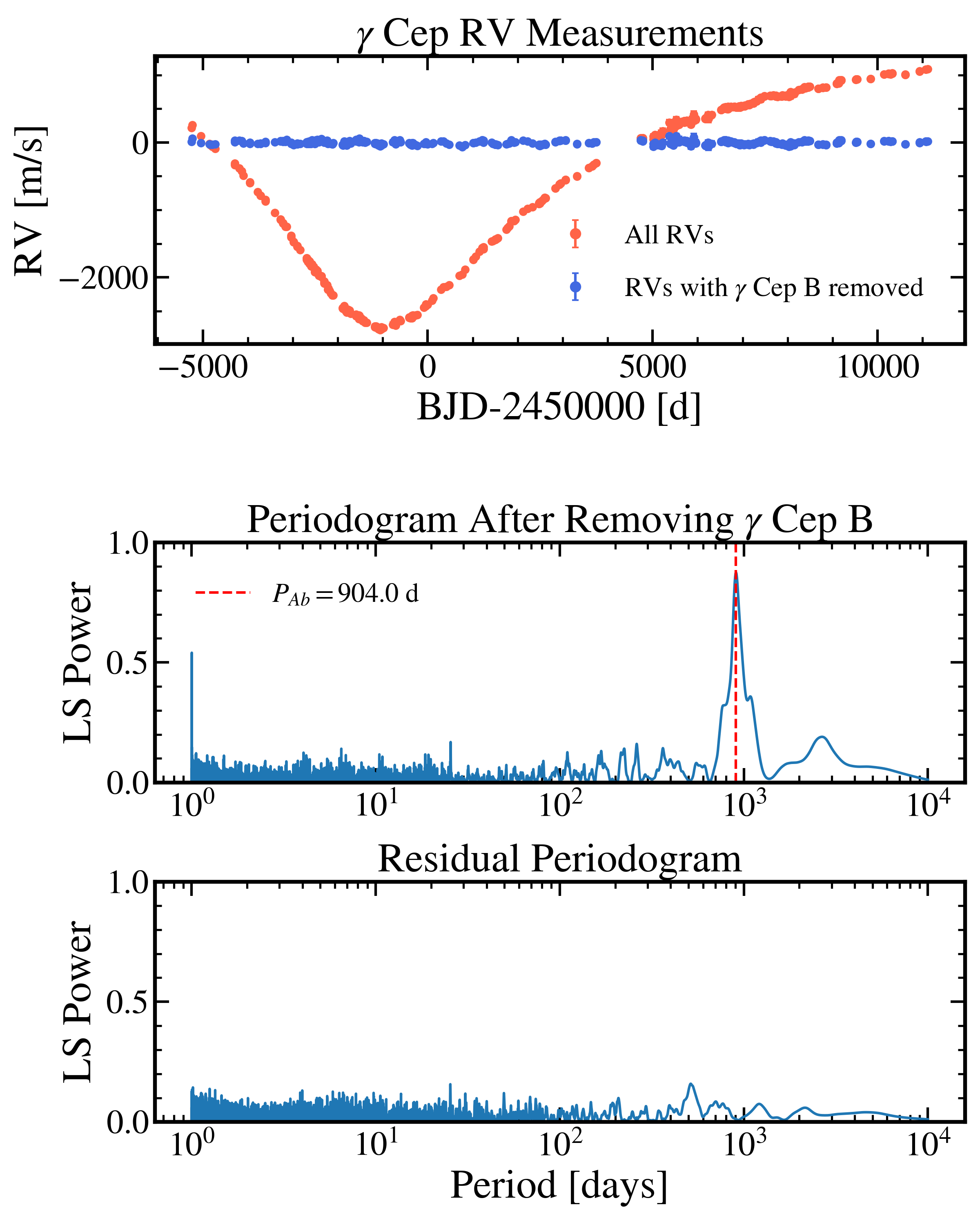}
    \caption{Periodogram search for additional signals in the RV timeseries of $\gamma$ Cep. The top panel shows the original RVs (orange) and the residuals after subtracting our best-fit model for $\gamma$ Cep B (blue). The middle panel shows the Lomb Scargle periodogram calculated from the RVs after removing $\gamma$ Cep B's signal. There is a clear peak at the period of $\gamma$ Cep Ab. The bottom panel shows the periodogram for the residuals after removing both $\gamma$ Cep B and Ab. No significant peaks are apparent.}
    \label{fig:periodogram}
\end{figure*}

We performed an additional search for periodic signals using a generalized Lomb Scargle (LS) periodogram \citep{Lomb1976, Scargle1982}. The periodogram search is more cursory than a full suite of injection/recovery tests in that it does not account for our search algorithm's sensitivity to periodic signals. However, it is also less computationally expensive, permitting the exploration of a wider range of orbital periods. We calculated the LS power over a fine period grid between 1 and $10^4$ days. Because our RV dataset does not cover a full period of $\gamma$ Cep B, we subtracted this signal from the RVs before computing the periodogram. We found strong power at $\PAb$ years, the period of $\gamma$ Cep Ab. We then subtracted $\gamma$ Cep Ab's signature from the RVs and computed a second periodogram to search for undetected short-period signals. We found no significant power in the residuals. Our periodogram analysis is shown in Figure \ref{fig:periodogram}.

\section{Discussion}
\label{sec:discussion}

Beyond its extensive observational history and status as one of the first candidate extrasolar planet hosts, $\gamma$ Cephei is an intriguing system due to its rich dynamical structure. The observations and analysis we presented in this study support the picture that has gradually materialized over the last 40 years: the evolved K-type RGB star $\gamma$ Cep A is orbited by an eccentric ($e\sim0.4$) M dwarf companion at $\sim20$ AU, as well as a Jovian planet on an S-type orbit --- that is, orbiting the primary star in a binary system --- at $\sim2$ AU.

Despite the general agreement reached on these points, outstanding questions about this system remain, such as how 
$\gamma$ Cep B and Ab arrived at their present configuration, and how $\gamma$ Cep Ab has remained stable in the presence of such a close stellar companion.

\subsection{$\gamma$ Cep Ab's formation}

The dynamical influence of a nearby star may pose a major obstacle to planet formation by truncating the protoplanetary disk and disrupting the coagulation of planetesimals needed for the formation of planetary cores \citep{ArtymowiczLubow1994, KleyBurkert2000, Ji2014, Kraus2016}. Nevertheless, over 200 planets have been discovered in binary systems, and over 50 have been discovered in multiple star systems \citep{Schwarz2016}.\footnote{Catalog of Exoplanets in Binary Star Systems: https://adg.univie.ac.at/schwarz/multiple.html} With a binary separation of $\sim20$ AU, $\gamma$ Cep is one of the closest-separation systems hosting an S-type planet \citep{ThebaultBonanni2025}, and the comparatively wide separation of $\gamma$ Cep Ab (1.98 AU) gives this system a planet-to-binary separation ratio of $\sim10\%$, the second-largest, after HD~7449 b (13\%), among the sample of 110 RV-detected S-type planets analyzed by \cite{Su2021}. Multiple mechanisms have been proposed to explain the origin of this architecture \citep{Thebault2011}, including large initial planetesimals that could resist perturbations from the companion \citep{Morbidelli2009}, an early ``snowball" phase during which planetesimals grow by dust accretion until they are massive enough to resist the companion's perturbations \citep{Xie2010}, disk instability as an alternative to core accretion \citep{Duchene2010}, and a wide primordial binary separation that later diminished due to interactions with a wide-separation or unbound third star \citep{Malmberg2007}.

The last of these was first explored by \cite{MarzariBarbieri2007}, who simulated a hierarchical triple system consisting of a stellar binary ($a_{in}=70$ AU) with a circumprimary disk, orbited by an inclined distant star ($a_{out}=212$ AU, $i_{mutual}=90^{\circ}$). The system underwent von Zeipel-Lidov-Kozai (vZLK) oscillations \citep{Kozai1962, Lidov1962, Ito2019} on a time scale of $\sim10^4$ years, destabilizing the disk due to the high eccentricities reached by the binary orbit. Under this picture, disk instability \citep{Boss1997} better explains the formation of $\gamma$ Cep Ab because, unlike core accretion, this mechanism would have sufficient time to generate the planet before the disk was disrupted. Encounters with an unbound star were later simulated specifically for the $\gamma$ Cep system by \cite{MartiBeauge2012}, who found that stellar fly-bys were able to reproduce the system's architecture, but only in 1--5\% of simulations. This relatively low probability may explain why S-type planets in close binaries are much rarer than in wide binary systems.

Another possibility is that a binary companion may actually aid certain planet formation pathways \citep{Zhang2018}. In this picture, an inclined binary companion at separations between $40-200$ AU could catalyze planet growth near 1 AU by exciting planetesimal collisions, resulting in lower planet multiplicity and higher planet masses. The consistency of this scenario with the $\gamma$ Cep system is unclear. On the one hand, our sensitivity analysis suggests that there are no undetected Jovians ($M\sin i \gtrsim 130~M_{\oplus}$) in the system. Further, $\gamma$ Cep Ab's minimum mass is greater than that of Jupiter, and a precise inclination measurement may reveal an even greater true mass. On the other hand, $\gamma$ Cep B's $20$ AU separation is significantly closer than the ``intermediate" range of $40-200$ AU quoted by \citep{Zhang2018}, and $\gamma$ Cep Ab's $1.98$ AU separation is likewise too wide to fit their picture of enhanced planet formation near 1 AU.

The relatively small number of close binaries ($<30$ AU) hosting S-type planets (e.g. Gl 86, \citealt{Zeng2022}; HD 7449, \citealt{Dumusque2011}; HD 196885, \citealt{Correia2008}) may indicate that the processes governing their formation are highly specialized, potentially involving different combinations of the pathways discussed here or other pathways entirely. In that case, system-specific analyses (e.g. \citealt{KleyNelson2008, Giuppone2011, MartiBeauge2012, Jordan2021}) are crucial to understanding systems like $\gamma$ Cep.

\subsection{$\gamma$ Cep's mutual inclination and long-term stability}

$\gamma$ Cep B's eccentric orbit brings it within $a(1-e)=20\text{ AU }\times(1-0.42) \approx 11.6$ AU of the primary star, a distance at which gravitational perturbations to $\gamma$ Cep Ab might be significant. Using the empirical stability criterion for S-type planets in binary systems calculated by \cite{HolmanWiegert1999}, \cite{Haghighipour2006} evaluated $\gamma$ Cep Ab's stability for a range of binary eccentricities $e_B$ and mutual inclinations $i_{\text{mut}}$. They found that for flat systems ($i_{\text{mut}}\sim0^{\circ}$) and at eccentricities matching our measured value of $e_B=0.42$, stable planetary orbits exist only within $\sim$2.6 AU, just outside $\gamma$ Cep Ab's measured separation of 1.98 AU. In the coplanar scenario, the planetary eccentricity is restricted to $e_{Ab}\lessapprox0.32$ by this stability boundary, because higher eccentricities would take the planet into the instability region at apoastron.

Furthermore, there is reason to believe that the system is misaligned ($i_{\text{mut}}>0^{\circ}$). \cite{ReffertQuirrenbach2011} used the re-reduction of the $Hipparcos$ catalog by \cite{vanLeeuwen2007} to constrain $\gamma$ Cep Ab's inclination. They found two local minima in their $\chi^2$ distribution of $i_{Ab}$, corresponding to $3\sigma$ confidence intervals of $3.8^{\circ}$--$20.8^{\circ}$ and $166.6^{\circ}$--$174.8^{\circ}$. \cite{Benedict2018} employed independent absolute astrometry from the Hubble Space Telescope Fine Guidance Sensor to measure $i_{Ab}=169.5^{\circ} \pm 1.1^{\circ}$.

Assessing the reliability of these measurements is difficult because the two aforementioned studies present the only measurements of $i_{Ab}$ of which we are aware. To make a more robust comparison, we analyze available measurements of the inclinations $i$, longitudes of the ascending node $\Omega$, and arguments of periastron $\omega$ of both the planetary and stellar companions (see Table \ref{tab:inc_Om_om_comparison}). In short, our measurements of these parameters for $\gamma$ Cep B agree well with past studies, with two exceptions. Our measurement of $\omega_{Ab}=\oAb^{\circ}$ is in marginal tension with the value of $243^{\circ} \pm 27^{\circ}$ obtained by both \cite{Torres07} and \cite{Neuhauser07}. To ascertain the source of this disagreement, we refit our orbital model with the subset of our precision RVs acquired before 1 January 2007, finding $\omega_{Ab}=260^{\circ} \pm 30^{\circ}$. This better agreement with \cite{Torres07} and \cite{Neuhauser07} suggests that the RVs collected over the last 20 years are the source of the discrepancy, and have improved the constraints on the orbit of $\gamma$ Cep Ab.

We found a more notable disagreement with the $i_B$ measurement of \cite{Benedict2018}, whose reported value of $101.5^{\circ} \pm 1.5^{\circ}$ is $\sim$12$\sigma$ from the more commonly found values of 118$^{\circ}$--120$^{\circ}$. This study's $\Omega_B$ measurement is likewise discrepant with the median value found among the other studies in Table \ref{tab:inc_Om_om_comparison} at $>4\sigma$. Thus, despite the agreement between their $i_{Ab}$ measurement and that of \cite{ReffertQuirrenbach2011}, it is unclear whether \cite{Benedict2018}'s inclination values are reliable. Additionally, these two studies show disagreement in $\Omega_{Ab}$: \cite{Benedict2018}'s value of $47^{\circ} \pm 6^{\circ}$ is separated by $\sim3\sigma$ from \cite{ReffertQuirrenbach2011}'s interval of $330.2^{\circ}$--$25.0^{\circ}$. 

In light of these collective inconsistencies, we conclude that decisive measurements of $i_{Ab}$ and $\Omega_{Ab}$ have not yet been made. Nevertheless, a high mutual inclination between $\gamma$ Cep B and Ab remains plausible, and a discussion of system stability in that scenario is therefore warranted.

\begin{deluxetable*}{lcccccc}
\tablecaption{Literature Values of $\gamma$ Cep's Angular Orbital Parameters}
\label{tab:inc_Om_om_comparison}
\tablehead{
\colhead{} &
\colhead{$i_B$} &
\colhead{$\Omega_B$} &
\colhead{$\omega_B$\tablenotemark{a}} &
\colhead{$i_{Ab}$} &
\colhead{$\Omega_{Ab}$} &
\colhead{$\omega_{Ab}$}\\
\colhead{Source} &
\colhead{(deg)} &
\colhead{(deg)} &
\colhead{(deg)} &
\colhead{(deg)} &
\colhead{(deg)} &
\colhead{(deg)}
}
\startdata
\cite{Torres07} & $118.1 \pm 1.2$ & $13.0 \pm 2.4$ & $340.96 \pm 0.40$ & --- & --- & $243 \pm 27$ \\
\cite{Neuhauser07} & $119.3 \pm 1.0$ & $18.04 \pm 0.98$ & $341.01 \pm 0.40$ &  --- & --- & $243 \pm 27$ \\
\cite{ReffertQuirrenbach2011}\tablenotemark{b} & --- & --- & --- & \makecell{3.8--20.8 \\ 166.6--174.8} & \makecell{352.9--86.0 \\ 330.2--25.0} & --- \\
\cite{Benedict2018} & $101.5 \pm 1.5$ & $14.2 \pm 0.8$ & $341.9 \pm 0.2$ & $169.5 \pm 1.1$ & $47 \pm 6$ & --- \\
\cite{Mugrauer22} & $120.18 \pm 0.27$ & $18.32 \pm 0.78$ & $340.49 \pm 0.50$ & --- & --- & --- \\
This work & \iB & \OB & \oB & --- & --- & \oAb
\enddata
\tablenotetext{a}{\cite{Torres07} and \cite{Neuhauser07} reported the argument of periastron of the host star's orbit. We calculated $\omega_{B}$ as $\omega_A$+$180^{\circ}$ and did the same for $\omega_{Ab}$.}
\tablenotetext{b}{\cite{ReffertQuirrenbach2011} reported multiple modes from their $\chi^2$ surface. Ranges represent $3\sigma$ confidence intervals. Intervals on the same line correspond to the same model.}
\end{deluxetable*}

In addition to flat configurations, \cite{Haghighipour2006} simulated $\gamma$ Cep's dynamical behavior over a sparse grid of mutual inclinations, $i_{\text{mut}}$=[0, 2$^{\circ}$, 5$^{\circ}$, 10$^{\circ}$, 20$^{\circ}$, 40$^{\circ}$, 60$^{\circ}$, 80$^{\circ}$], where $i_{\text{mut}}$ is related to the companion inclinations by:

\begin{equation}
\begin{aligned}
\cos(i_{\mathrm{mut}}) &= \cos(i_B)\cos(i_{Ab}) \\
&{}+ \sin(i_B)\sin(i_{Ab}) \cos(\Omega_B - \Omega_{Ab}).
\end{aligned}
\end{equation}

\cite{Haghighipour2006} found that the system was unstable for $i_{\text{mut}}>40^{\circ}$ except when $i_{\text{mut}}=60^{\circ}$, in which case the system underwent stable eccentric vZLK oscillations (EKL; \citealt{LithwickNaoz2011, Naoz2016}). \cite{HuangJi2022} revisited the EKL mechanism in this system, assuming \cite{ReffertQuirrenbach2011}'s best-fit planetary parameters --- $i_{Ab}=5.7^{\circ}$, $\Omega_{Ab}=37.5^{\circ}$ --- and those of \cite{Neuhauser07} for the binary analogs --- $i_{B}=119.3^{\circ}$, $\Omega_{B}=18.04^{\circ}$ --- which yield $i_{\text{mut}}=113.9^{\circ}$. They determined that a true mass of $M_{Ab}$=15 $\Mjup$ would allow $\gamma$ Cep to reach this mutual inclination within the system lifetime, provided that the initial mutual inclination and planetary eccentricity satisfy $i_{\text{mut}}<60^{\circ}$, $e_{Ab}<0.7$. The alternative $i_{\text{mut}}$ of $54.3^{\circ}$, resulting from \cite{ReffertQuirrenbach2011}'s second solution of $i_{Ab}=173.1^{\circ}$, $\Omega_{Ab}=356.1^{\circ}$, could be attained under a wider range of planetary masses and initial conditions because it does not require an orbit flip, assuming $i_{\text{mut}}$ was initially $<90^{\circ}$. Thus, both of \cite{ReffertQuirrenbach2011}'s solutions for the orbit of $\gamma$ Cep Ab correspond to EKL oscillations that maintain long-term stability in the system.

\subsection{Testing M dwarf Mass-Magnitude Relations}

The dynamical mass of $\MBsolar~\Msun$ we derive for $\gamma$ Cep B has an uncertainty of $0.5\%$, among the most precise M dwarf mass measurements made (e.g., \citealt{Benedict2016, Dupuy2017, Mann2019}). Here, we use this measurement to test the M dwarf mass-magnitude relations derived by \cite{Benedict2016} and \cite{Mann2019}. We began with the measurement of $\gamma$ Cep B's apparent $K$-band magnitude, $K=7.3 \pm 0.2$ mag, made by \cite{Neuhauser07}, from which we derive an absolute magnitude of $M_K=6.6\pm0.2$ mag. The polynomial fit of \cite{Benedict2016}, derived for stars between $0.08-0.62~\Msun$ from the dynamical masses of 47 M dwarfs, produces an estimate of $0.38\pm0.04~\Msun$, fully consistent with our fitted value. 

Meanwhile, the relation derived by \cite{Mann2019} for stars between $0.075-0.7~\Msun$ from 62 binary systems requires the $K_s$ magnitude as an input. We use $\gamma$ Cep B's mass to estimate a spectral type of M2.5V \citep{Pecaut&Mamajek2013} and retrieved the spectrum of a similar M dwarf, Gl 581, from the IRTF Spectral Library \citep{Rayner2009}. Synthetic photometry of this spectrum yields a $K_s-K$ color of $0.03$, which is within the uncertainty of $\gamma$ Cep B's $K$-band magnitude. We use the \cite{Mann2019} relation with our synthetic $K_s$-band magnitude to estimate $M_B=0.34\pm0.03~\Msun$, in mild tension with both our fitted mass ($1.9\sigma$) and that modeled from the \cite{Benedict2016} relation ($1.3\sigma$).

\section{Conclusion}
\label{sec:conclusion}

Our observations represent the latest chapter in the decades-long study of the $\gamma$ Cephei system. We derived updated orbital parameters using both new and published datasets. Our new observations include four RV datasets of $\gamma$ Cep A from Keck/HIRES, APF/Levy, McDonald/TS23, and Whipple/TRES, and eight epochs of relative astrometry from Keck/NIRC2. Our augmented dataset allowed us to derive precise constraints on the masses and orbital parameters of $\gamma$ Cep B and Ab, improving precision by a factor of 2--10 compared to previous work. 

Our results confirm that $\gamma$ Cep B is an early M dwarf ($M_B=\MBsolar \, \Msun$) with a period of $\PB$ years, and that $\gamma$ Cep Ab is a super-Jupiter ($M \sin i$=$\MAbsini \, \Mjup$) with a period of $\PAb$ years. Our findings are compatible with the conclusions of past work that the EKL mechanism may play a strong part in maintaining the stability of this system, which would otherwise be tenuous due to $\gamma$ Cep B's eccentricity and relatively narrow separation from $\gamma$ Cep Ab.

The upcoming publication of the $Gaia$ mission's fourth data release (DR4) will offer improved constraints on the orientation of $\gamma$ Cep Ab's orbit, the last undetermined component of this system. Although $\gamma$ Cep A's brightness ($G$=2.9) places it just outside of $Gaia$'s nominal bright limit of $G$=3, the star's DR3 solution ($Gaia$ DR3 2281778105594488192) suggests that it will appear in the DR4 catalog. DR4's enhanced precision should reduce the uncertainty on $\gamma$ Cep's parallax by a factor of 10 \citep{Brown2025}, and its 66-month baseline will provide epoch astrometry over more than two full orbits of the planetary companion. Moreover, the fact that $Gaia$'s mission window is contemporaneous with many of the precision RVs we have presented here will maximize the utility of a three-dimensional joint orbital fit. Measuring the inclination of $\gamma$ Cep Ab with high precision is key to determining its true mass and $i_{\text{mut}}$, both of which will grant insight into the long-term stability and dynamical evolution of this system.

\section{Acknowledgments}
This paper includes data taken at The McDonald Observatory of The University of Texas at Austin.
J.W.X is grateful for support from the Heising-Simons Foundation 51 Pegasi b Fellowship (grant \#2025-5887). 

\software{\texttt{astropy} \citep{astropy:2013, astropy:2018, astropy:2022},
          \texttt{ethraid} \citep{VanZandt2024},
          \texttt{numpy} \citep{numpy:2020}, 
          \texttt{scipy} \citep{scipy:2020},
          }

\bibliography{gamma_cep}
\bibliographystyle{aasjournal}

\end{document}